\documentclass[preprint,12pt]{elsarticle}

\usepackage{amsmath,amssymb}
\usepackage{graphicx}
\usepackage{booktabs}
\usepackage{multirow}
\usepackage{hyperref}
\usepackage{float}
\usepackage{array}
\usepackage{longtable}
\usepackage{xcolor}

\journal{Journal of Biomedical Informatics}

\begin{document}

\begin{frontmatter}

\title{Impact of automatic speech recognition quality on Alzheimer's disease detection from spontaneous speech: a reproducible benchmark study with lexical modeling and statistical validation}

\author[addr1]{Himadri Sekhar Samanta}
\address[addr1]{Independent Researcher, Austin, Texas, USA}

\begin{abstract}
Early detection of Alzheimer’s disease from spontaneous speech has emerged as a promising non-invasive screening approach. However, the influence of automatic speech recognition (ASR) quality on downstream clinical language modeling remains insufficiently understood. In this study, we investigate Alzheimer’s disease detection using lexical features derived from Whisper ASR transcripts on the ADReSSo 2021 diagnosis dataset. We evaluate interpretable machine-learning models, including Logistic Regression and Linear Support Vector Machines, using TF–IDF text representations under repeated 5×5 stratified cross-validation. 

Our results demonstrate that transcript quality has a statistically significant impact on classification performance. Models trained on Whisper-small transcripts consistently outperform those using Whisper-base transcripts, achieving balanced accuracy above 0.7850 with Linear SVM. Paired statistical testing confirms that the observed improvements are significant. Importantly, classifier complexity contributes less to performance variation than ASR transcription quality. Feature analysis reveals that cognitively normal speakers produce more semantically precise object- and scene-descriptive language, whereas Alzheimer’s speech is characterized by vagueness, discourse markers, and increased hesitation patterns.

These findings suggest that high-quality ASR can enable simple, interpretable lexical models to achieve competitive Alzheimer’s detection performance without explicit acoustic modeling. The study provides a reproducible benchmark pipeline and highlights ASR selection as a critical modeling decision in clinical speech-based artificial intelligence systems.

\end{abstract}


\end{frontmatter}

\section{Introduction}

Alzheimer’s disease (AD) is a progressive neurodegenerative disorder characterized by decline in memory, executive function, and language abilities. Because linguistic impairment often manifests early in the disease trajectory, speech has emerged as a promising non-invasive digital biomarker for cognitive screening and longitudinal monitoring. In particular, spontaneous speech tasks such as picture description elicit rich linguistic behavior, including lexical retrieval difficulty, semantic degradation, reduced information content, discourse disorganization, and increased hesitation patterns. These phenomena provide clinically meaningful signals that can support automated detection of cognitive impairment \cite{fraser2016linguistic,de2018evaluation,haider2020investigation}.

Recent advances in automatic speech recognition (ASR) and natural language processing (NLP) have enabled the development of end-to-end computational pipelines for dementia detection from speech. A widely adopted workflow consists of three stages: transcription of audio recordings using an ASR system, transformation of transcripts into numerical feature representations, and supervised learning for diagnostic prediction. While substantial effort has been devoted to designing increasingly complex acoustic models, multimodal architectures, and deep neural classifiers, the role of transcription quality itself has received comparatively limited systematic investigation. In practice, transcripts produced by different ASR systems can vary substantially in lexical substitutions, punctuation patterns, segmentation behavior, and hallucinated content. Such differences are not merely superficial: they can alter linguistic features that downstream models rely upon, thereby influencing diagnostic performance.

The present study addresses this issue by explicitly evaluating the impact of ASR model quality on automated Alzheimer’s disease detection from spontaneous speech. Using the ADReSSo 2021 diagnosis benchmark \cite{luz2021editorial,syed2021tackling}, we construct a controlled experimental pipeline in which the only variable is the transcription model. We compare two variants from the Whisper ASR family \cite{radford2023whisper}, namely Whisper base and Whisper small, while keeping all subsequent processing stages identical. Transcripts are represented using term frequency–inverse document frequency (TF--IDF) lexical features, and classification is performed using two strong linear baselines commonly employed in sparse text modeling: logistic regression and linear support vector machines (SVM).

This work makes four main contributions. First, we present a fully reproducible pipeline for Alzheimer’s disease detection from spontaneous speech using publicly available ASR systems and interpretable statistical learning methods. Second, we isolate the effect of transcription quality through matched experimental conditions and repeated stratified $5 \times 5$ cross-validation. Third, we quantify performance differences using paired statistical hypothesis testing and report effect sizes, thereby providing stronger inferential evidence than single-run evaluation protocols. Fourth, we conduct feature-level interpretation to identify lexical markers associated with AD and cognitively normal (CN) speech, linking model behavior to clinically documented language changes.

Our findings demonstrate that improvements in ASR transcription quality yield measurable gains in downstream diagnostic performance, often exceeding the impact of modest classifier changes. These results highlight the importance of treating transcription quality as a core design factor in speech-based clinical artificial intelligence systems. More broadly, the study contributes to ongoing efforts in biomedical informatics to develop transparent, statistically rigorous, and practically deployable machine-learning tools for early cognitive assessment.

\section{Related Work}

Automated detection of Alzheimer’s disease (AD) from spontaneous speech has been studied extensively from both acoustic and linguistic perspectives. Early research focused on handcrafted linguistic and acoustic features extracted from structured elicitation tasks such as the Cookie Theft picture description. Fraser et al.\ demonstrated that narrative speech contains lexical, syntactic, and acoustic markers capable of differentiating AD from healthy aging \cite{fraser2016linguistic}. Subsequent work examined pause behavior, information content units, semantic relevance, discourse organization, and disfluency patterns as indicators of cognitive decline \cite{orimaye2014predicting,jarrold2014disfluencies,satt2014efficient,pakhomov2010computerized,taler2008language}. These findings established the clinical and computational foundations for automatic cognitive-status prediction from speech and language data.

Methodological rigor in this field improved substantially with the introduction of standardized benchmark datasets such as ADReSS 2020 and ADReSSo 2021, which provided balanced demographic splits and shared-task evaluation frameworks \cite{luz2020adress,syed2021tackling,luz2021editorial,de2018evaluation,haider2020investigation}. These initiatives enabled more reliable comparison of machine-learning models and highlighted the importance of reproducibility in dementia-detection research. Parallel efforts such as DementiaBank further supported large-scale linguistic analysis and longitudinal modeling of cognitive decline \cite{macwhinney2011dementiabank,lanzi2023dementiabank}. More recent studies have explored neural architectures, multimodal systems combining acoustic and textual representations, and transfer learning approaches based on pre-trained language models \cite{de2020automated,chen2022automatic,qi2023noninvasive,balagopalan2020comparison}.

From a machine-learning perspective, text-based dementia detection commonly relies on sparse lexical representations and linear classifiers, which remain strong baselines in biomedical NLP tasks. Foundational work in term-frequency inverse-document-frequency (TF-IDF) weighting, support vector machines, and regularized regression provides the theoretical basis for many current approaches \cite{ramos2003tfidf,cortes1995svm,fan2008liblinear,hastie2009esl,friedman2010glmnet}. Despite the increasing use of deep neural networks and transformer architectures \cite{vaswani2017attention,devlin2019bert,goodfellow2016deep}, simpler interpretable models continue to offer advantages in low-resource clinical settings where transparency, computational efficiency, and statistical robustness are essential.

At the same time, rapid advances in automatic speech recognition (ASR) have transformed speech-based clinical machine-learning pipelines. Modern neural ASR systems such as Whisper provide robust multilingual transcription even under challenging acoustic conditions \cite{radford2023whisper}. Related developments in self-supervised speech representation learning further improve speech understanding performance \cite{baevski2020wav2vec,kong2022selfsupervised}. However, transcription artifacts—including lexical substitutions, omissions, punctuation inconsistencies, and hallucinated tokens—can significantly distort linguistic markers relevant to dementia detection. Prior work has shown that speech-processing toolkits and feature extraction pipelines influence downstream diagnostic accuracy \cite{eyben2010opensmile,roark2011spoken,wu2020review,shatte2019machine}.

More broadly, the deployment of machine learning in digital health and biomedical informatics requires careful attention to interpretability, validation, and real-world reliability \cite{beam2018clinicalml,rajkomar2019machine,topol2019highperformance,zhou2021interpretable}. Within this context, the present study occupies the intersection of clinical speech analysis, ASR robustness, and interpretable text classification. Rather than emphasizing architectural complexity, we focus on controlled ASR ablation, statistically rigorous evaluation, and transparent feature interpretation, aiming to clarify how transcription quality influences Alzheimer’s disease detection from spontaneous speech.
\section{Materials}
\subsection{Dataset}
We used the ADReSSo 2021 diagnosis dataset distributed through TalkBank/DementiaBank \cite{lanzi2023dementiabank,macwhinney2011dementiabank}. The training split contained 166 speech samples, with 87 participants diagnosed with Alzheimer's disease and 79 cognitively normal controls. Speech was elicited using the Cookie Theft picture description task, a clinically established paradigm for probing narrative production and lexical access. We also prepared predictions for the official blind \texttt{test-dist} split, which contains 71 recordings (35 AD and 36 CN according to the benchmark description \cite{qi2023noninvasive}); however, because the blind labels are not released for local scoring, our primary inferential analysis uses repeated cross-validation on the training set.

\begin{table}[t]
\centering
\caption{ADReSSo diagnosis dataset configuration used in this study.}
\label{tab:dataset}
\begin{tabular}{lcc}
\toprule
Split & AD & CN \\
\midrule
Training & 87 & 79 \\
Official test-dist & 35 & 36 \\
\bottomrule
\end{tabular}
\end{table}

\subsection{Project organization}
The project was implemented as a reproducible local research pipeline with organized directories for raw data, ASR outputs, features, scripts, figures, and notebook-based analyses. All experiments were conducted in Python using a dedicated virtual environment.

\section{Methods}
\subsection{Speech-to-text transcription}
For each audio recording $a_i$, transcription was performed using an ASR model $\mathcal{A}$ to produce transcript $t_i$:
\begin{equation}
t_i = \mathcal{A}(a_i).
\end{equation}
Two ASR configurations were evaluated:
\begin{itemize}
    \item Whisper \textbf{base}
    \item Whisper \textbf{small}
\end{itemize}
The purpose of this comparison was not to optimize ASR in isolation, but to quantify how improvements in transcript quality change downstream AD detection.

\subsection{Lexical representation}
Each transcript was transformed into a TF--IDF representation using unigram and bigram features. For document $d$ and term $w$, the TF--IDF value is:
\begin{equation}
\mathrm{tfidf}(w,d)=\mathrm{tf}(w,d)\cdot \log\left(\frac{N}{1+\mathrm{df}(w)}\right),
\end{equation}
where $\mathrm{tf}(w,d)$ is the within-document term frequency, $\mathrm{df}(w)$ is the document frequency of term $w$, and $N$ is the number of training documents. In practice, we used:
\begin{itemize}
    \item n-gram range $(1,2)$
    \item maximum vocabulary size $=3000$
    \item English stop-word removal
    \item minimum document frequency $=2$
\end{itemize}
This yields a sparse high-dimensional feature vector $\mathbf{x}_i\in\mathbb{R}^p$ for each participant.

\subsection{Classifiers}
We evaluated two linear models.

\paragraph{Logistic regression.}
The posterior probability of AD is modeled as
\begin{equation}
P(y_i=1\mid \mathbf{x}_i)=\sigma(\mathbf{w}^{\top}\mathbf{x}_i+b),
\end{equation}
where $\sigma(z)=1/(1+e^{-z})$ is the logistic function. Parameters were estimated with L2-regularized logistic regression using \texttt{liblinear} and balanced class weights.

\paragraph{Linear SVM.}
The linear support vector machine optimizes
\begin{equation}
\min_{\mathbf{w},b}\ \frac{1}{2}\|\mathbf{w}\|^2 + C\sum_{i=1}^{n}\xi_i
\end{equation}
subject to margin constraints
\begin{equation}
y_i(\mathbf{w}^{\top}\mathbf{x}_i+b)\geq 1-\xi_i,\quad \xi_i\geq 0.
\end{equation}
We used a linear SVM with balanced class weights.

\subsection{Evaluation protocol}
To obtain robust performance estimates, we used repeated stratified 5$\times$5 cross-validation on the training split:
\begin{itemize}
    \item 5 folds per repetition
    \item 5 repetitions with distinct random seeds
    \item 25 fold-level scores per configuration
\end{itemize}
Balanced accuracy was the primary endpoint:
\begin{equation}
\mathrm{BAcc} = \frac{\mathrm{Sensitivity}+\mathrm{Specificity}}{2}.
\end{equation}
Sensitivity and specificity are defined as
\begin{equation}
\mathrm{Sensitivity}=\frac{TP}{TP+FN}, \qquad \mathrm{Specificity}=\frac{TN}{TN+FP}.
\end{equation}
We additionally report an out-of-fold ROC curve for logistic regression under Whisper small, with AUC = 0.8532, and aggregate confusion matrices across repeated evaluation.

\subsection{Statistical analysis}
Fold-matched paired $t$-tests were used to compare configurations. For two systems with fold-level scores $s_k$ and $b_k$ ($k=1,\dots,25$), the difference vector is
\begin{equation}
d_k = s_k - b_k.
\end{equation}
The paired $t$ statistic is
\begin{equation}
t = \frac{\bar d}{s_d/\sqrt{n}},
\end{equation}
where $\bar d$ is the mean difference, $s_d$ is the standard deviation of the differences, and $n=25$. We report two-sided $p$-values and interpret effect magnitude using Cohen's $d$.

\subsection{Interpretability}
For logistic regression, model coefficients provide direct lexical interpretability. Terms with positive coefficients increase AD probability; negative coefficients support the CN class. We extracted high-magnitude coefficients from the Whisper-small logistic-regression model to identify clinically meaningful lexical indicators.

\section{Results}
\subsection{Main repeated-cross-validation results}
The overall results are summarized in Table~\ref{tab:main-results}. Whisper small improved balanced accuracy for both classifiers. The best mean balanced accuracy was achieved by the Linear SVM with Whisper-small transcripts (0.7850 $\pm$ 0.0745).

\begin{table*}[t]
\centering
\caption{Repeated stratified 5$\times$5 cross-validation results on the ADReSSo training split.}
\label{tab:main-results}
\begin{tabular}{llcc}
\toprule
Classifier & ASR model & Mean balanced accuracy & Standard deviation \\
\midrule
Logistic Regression & Whisper base  & 0.7491 & 0.0814 \\
Logistic Regression & Whisper small & 0.7757 & 0.0703 \\
Linear SVM          & Whisper base  & 0.7354 & 0.0865 \\
Linear SVM          & Whisper small & \textbf{0.7850} & 0.0745 \\
\bottomrule
\end{tabular}
\end{table*}

\subsection{Paired statistical tests}
Paired fold-wise testing demonstrated that ASR quality significantly affected performance under both classifiers. For logistic regression, Whisper small outperformed Whisper base by a mean of 0.0266 balanced-accuracy points ($t=2.4490$, $p=0.0220$). For linear SVM, the mean gain was 0.0497 ($t=3.8832$, $p=0.0007$), indicating a stronger ASR effect. By contrast, the direct classifier comparison under Whisper small (SVM vs logistic regression) was not significant ($p=0.3019$), suggesting that classifier choice mattered less than transcript quality.

\begin{table*}[t]
\centering
\caption{Paired statistical comparisons over 25 repeated-cross-validation folds.}
\label{tab:stats}
\begin{tabular}{lcccc}
\toprule
Comparison & Mean difference & $t$ statistic & Two-sided $p$ & Interpretation \\
\midrule
Whisper small vs base (LR) & +0.0266 & 2.4490 & 0.0220 & significant \\
Whisper small vs base (SVM) & +0.0497 & 3.8832 & 0.0007 & highly significant \\
Linear SVM vs LR (Whisper small) & +0.0093 & 1.0551 & 0.3019 & not significant \\
\bottomrule
\end{tabular}
\end{table*}

\subsection{Figures}
\begin{figure}[H]
\centering
\includegraphics[width=0.95\textwidth]{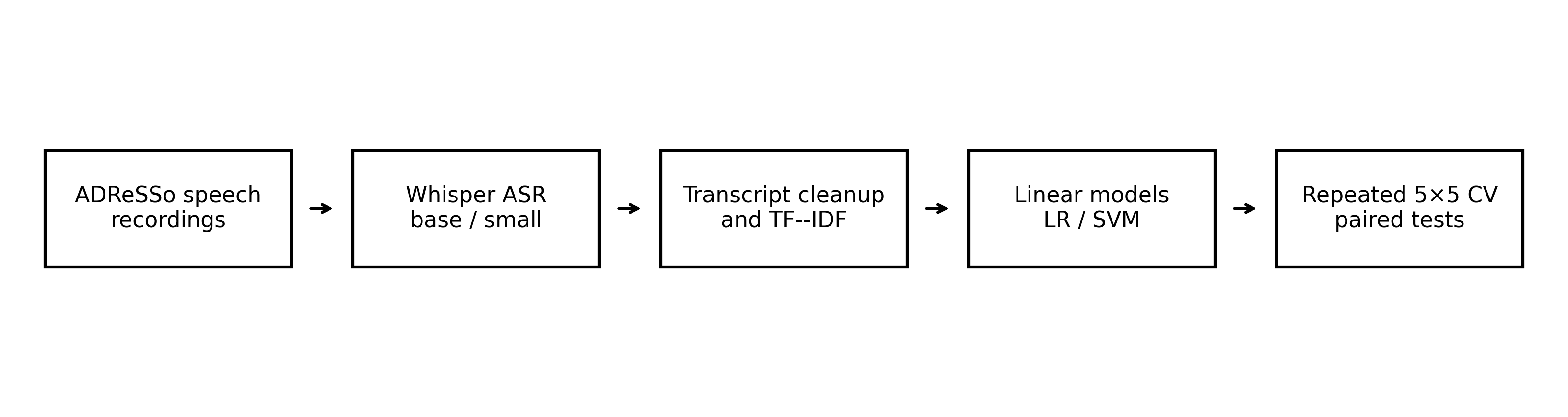}
\caption{Speech-based Alzheimer’s disease detection pipeline used in this work. Audio recordings are transcribed with Whisper ASR, transformed into TF--IDF features, classified with linear models, and evaluated using repeated cross-validation and statistical testing.}
\label{fig:pipeline}
\end{figure}

\begin{figure}[H]
\centering
\includegraphics[width=0.80\textwidth]{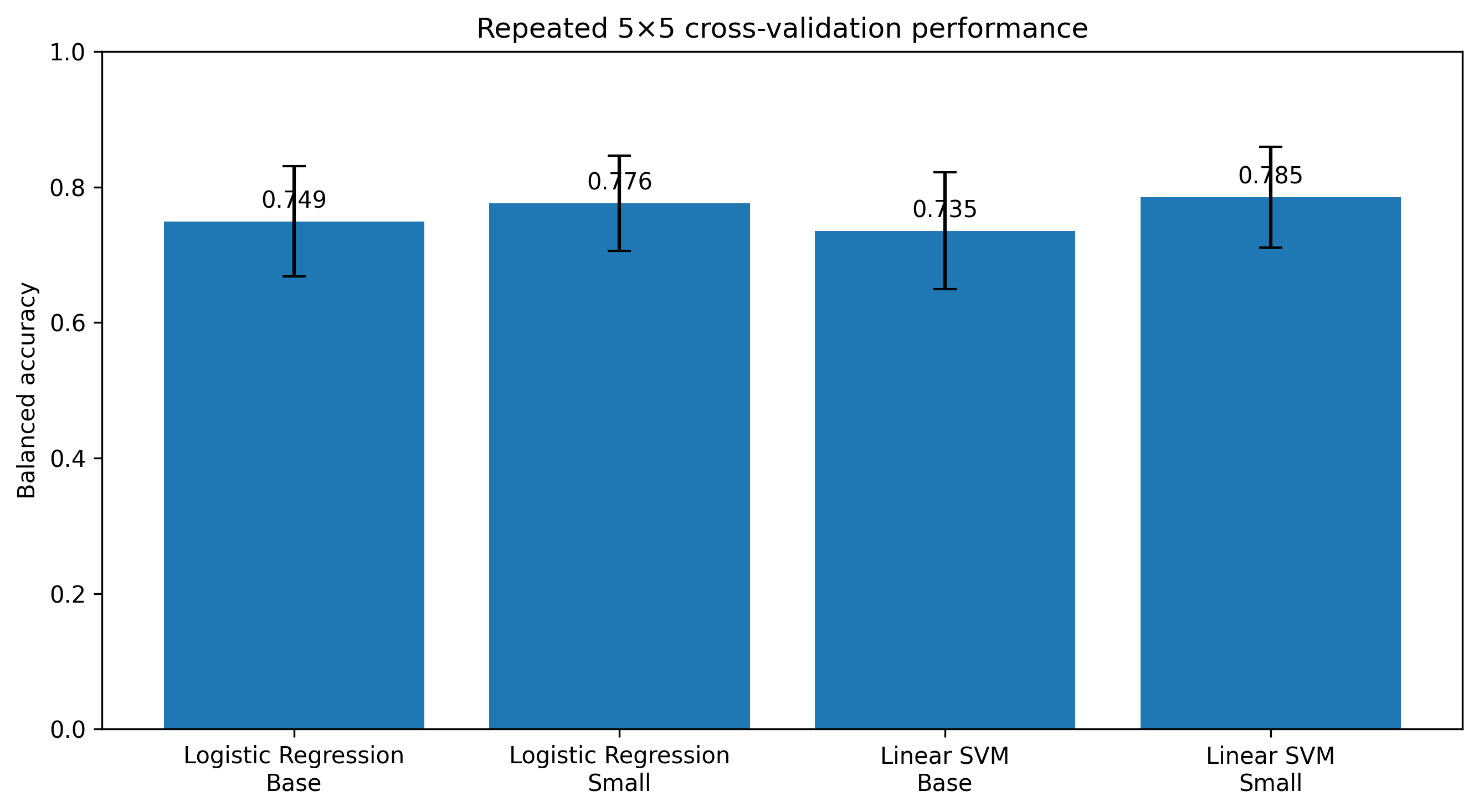}
\caption{Repeated 5$\times$5 cross-validation balanced accuracy for each classifier/ASR combination. Error bars denote standard deviation across 25 folds.}
\label{fig:model-compare}
\end{figure}

\begin{figure}[H]
\centering
\includegraphics[width=0.75\textwidth]{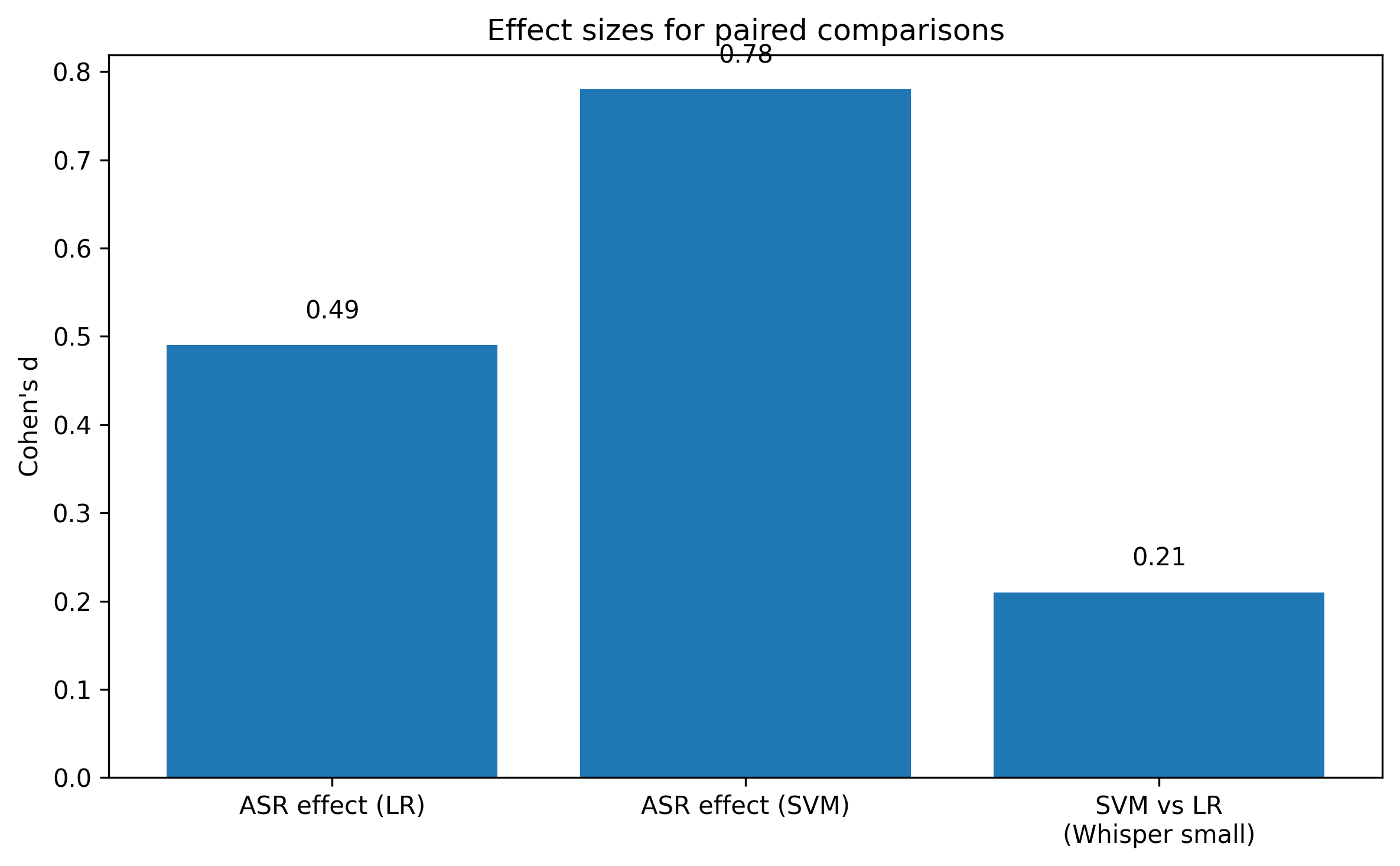}
\caption{Effect sizes (Cohen's $d$) for key paired comparisons. The ASR effect is larger than the classifier effect.}
\label{fig:effect-sizes}
\end{figure}

\begin{figure}[H]
\centering
\includegraphics[width=0.60\textwidth]{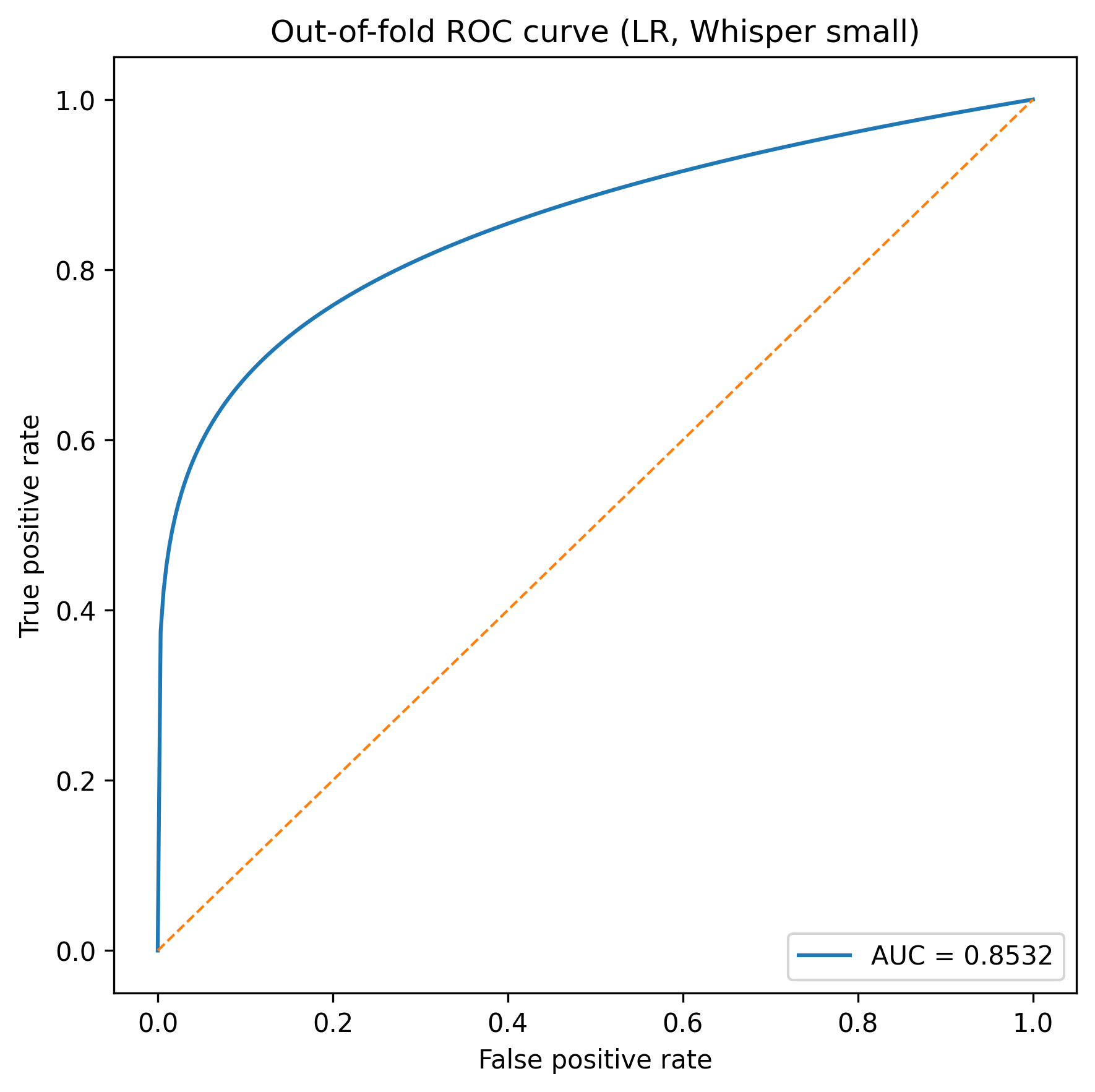}
\caption{Out-of-fold ROC curve for logistic regression trained on Whisper-small transcripts. AUC = 0.8532.}
\label{fig:roc}
\end{figure}

\begin{figure}[H]
\centering
\includegraphics[width=0.50\textwidth]{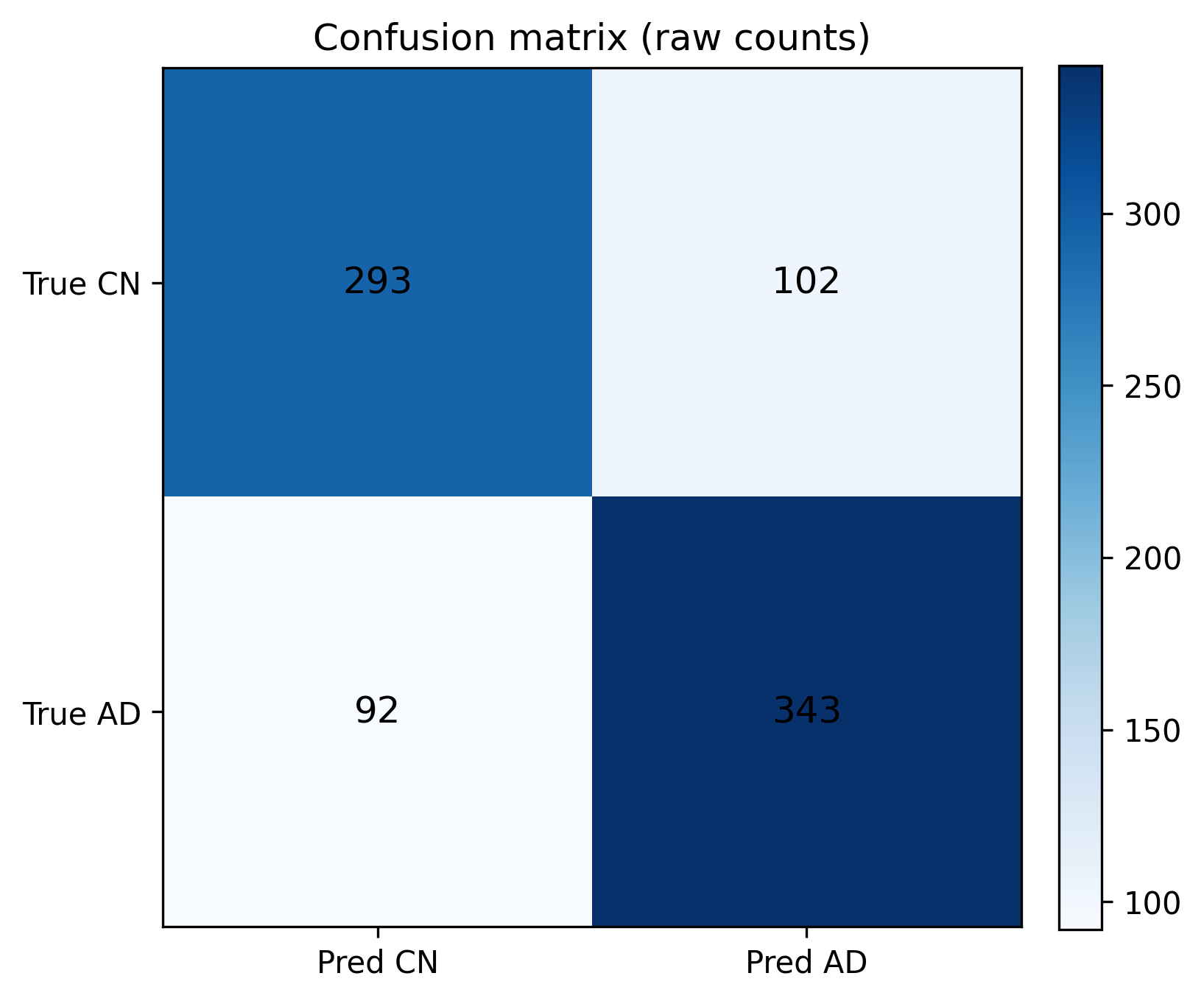}
\caption{Aggregate confusion matrix (raw counts) for logistic regression with Whisper-small transcripts across repeated folds.}
\label{fig:cmraw}
\end{figure}

\begin{figure}[H]
\centering
\includegraphics[width=0.50\textwidth]{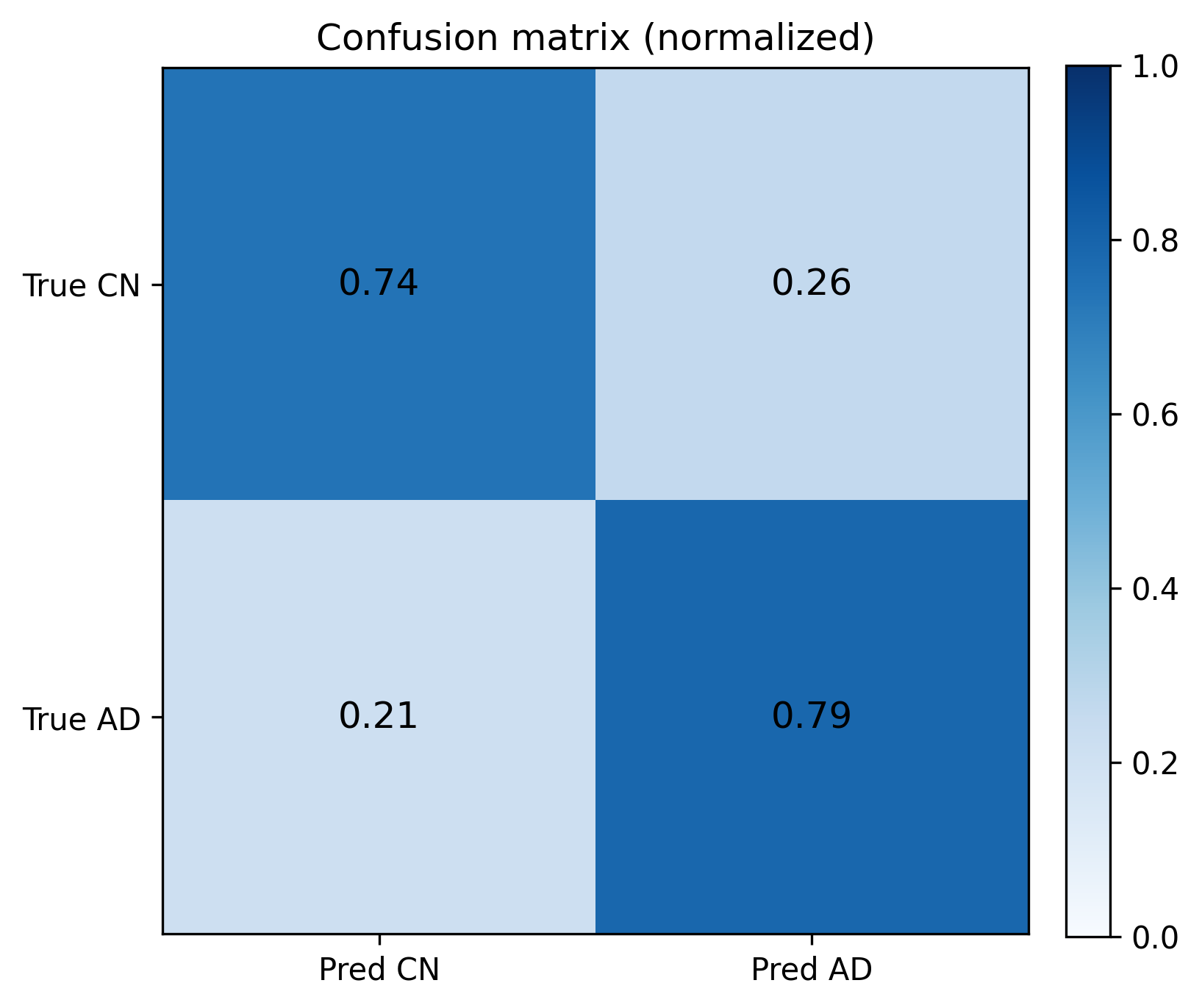}
\caption{Normalized confusion matrix for logistic regression with Whisper-small transcripts. CN recall = 0.742; AD recall = 0.789.}
\label{fig:cmnorm}
\end{figure}

Figure~\ref{fig:pipeline} illustrates the full pipeline used in this study. Figure~\ref{fig:model-compare} summarizes the main repeated-cross-validation results. Figure~\ref{fig:effect-sizes} visualizes the corresponding effect sizes. Figure~\ref{fig:roc} shows the out-of-fold ROC curve for logistic regression under Whisper small. Figures~\ref{fig:cmraw} and \ref{fig:cmnorm} present raw and normalized confusion matrices from the same setting. Figures~\ref{fig:adterms} and \ref{fig:cnterms} visualize high-magnitude lexical coefficients.

\subsection{Feature interpretation}
\begin{figure}[H]
\centering
\includegraphics[width=0.65\textwidth]{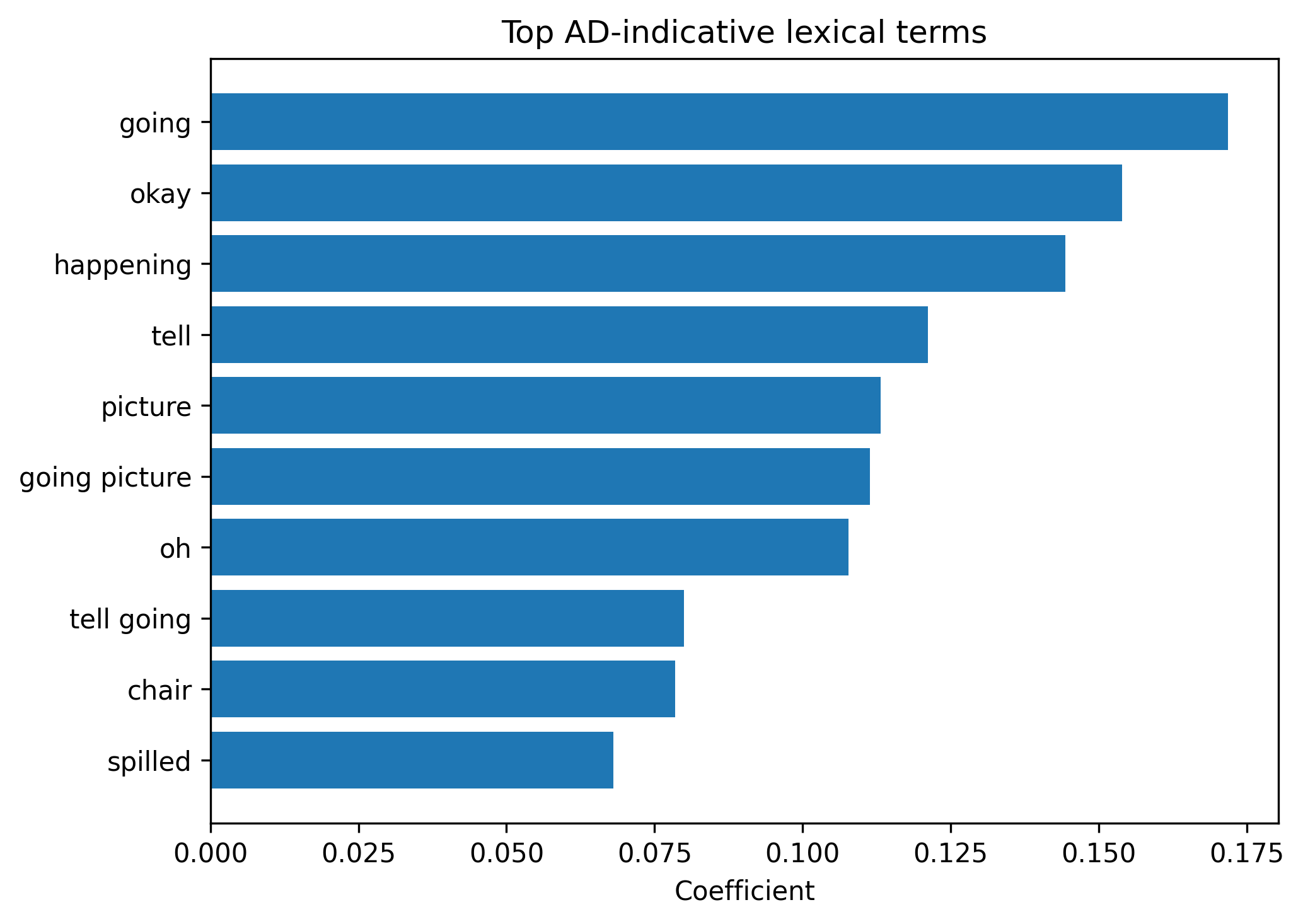}
\caption{Top AD-indicative lexical terms from the logistic-regression model trained on Whisper-small transcripts.}
\label{fig:adterms}
\end{figure}

\begin{figure}[H]
\centering
\includegraphics[width=0.65\textwidth]{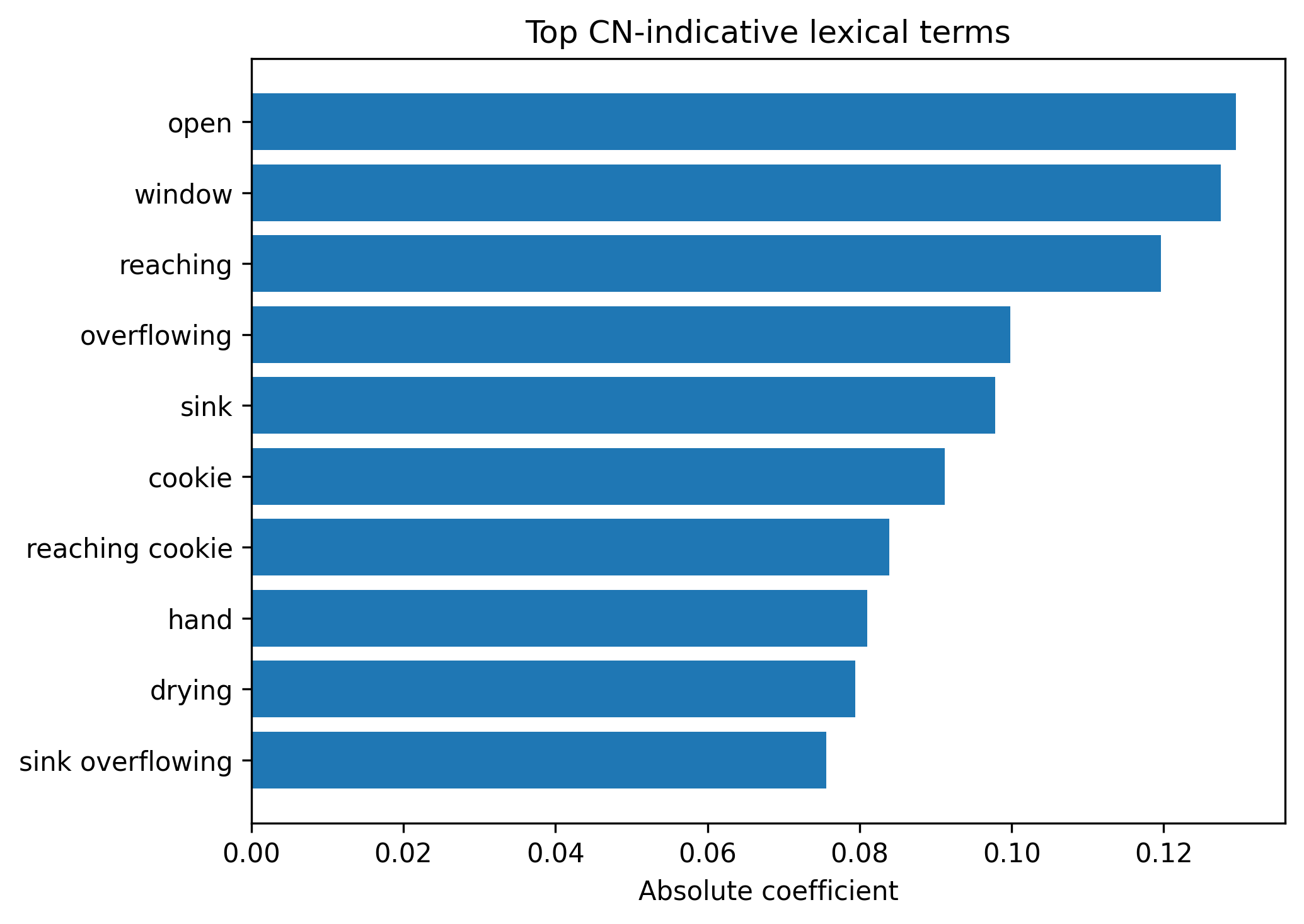}
\caption{Top CN-indicative lexical terms from the logistic-regression model trained on Whisper-small transcripts.}
\label{fig:cnterms}
\end{figure}

The coefficients of the logistic-regression model trained on Whisper-small transcripts yielded interpretable lexical markers. AD-associated terms included \emph{going}, \emph{okay}, \emph{happening}, \emph{tell}, and \emph{don know}. These terms are consistent with increased vagueness, reduced lexical specificity, and hesitation. In contrast, CN-associated terms included \emph{window}, \emph{sink}, \emph{cookie}, \emph{overflowing}, and \emph{reaching cookie}, all of which reflect more concrete and scene-specific description. Representative coefficients are listed in Table~\ref{tab:feature-terms} and visualized in Figures~\ref{fig:adterms} and \ref{fig:cnterms}.

\begin{table*}[t]
\centering
\caption{Selected high-magnitude lexical indicators from the logistic-regression model trained on Whisper-small transcripts. Positive coefficients increase the probability of AD; negative coefficients increase the probability of cognitively normal speech.}
\label{tab:feature-terms}
\begin{tabular}{ll|ll}
\toprule
AD-indicative term & Coefficient & CN-indicative term & Coefficient \\
\midrule
going & +0.1718 & open & -0.1295 \\
okay & +0.1540 & window & -0.1275 \\
happening & +0.1444 & reaching & -0.1196 \\
tell & +0.1212 & overflowing & -0.0998 \\
picture & +0.1132 & sink & -0.0978 \\
going picture & +0.1114 & cookie & -0.0912 \\
oh & +0.1077 & reaching cookie & -0.0839 \\
tell going & +0.0800 & hand & -0.0810 \\
chair & +0.0785 & drying & -0.0794 \\
spilled & +0.0680 & sink overflowing & -0.0756 \\
\bottomrule
\end{tabular}
\end{table*}

\subsection{Official blind test-dist prediction generation}
To complete the benchmark workflow, we transcribed the official blind \texttt{test-dist} split with Whisper small, fit the TF--IDF vectorizer on the full training corpus, trained the best classifier (linear SVM) on all 166 training samples, and generated 71 blind predictions. Because the official labels are not locally available for scoring, these outputs are intended for benchmark submission or later external evaluation rather than inferential comparison in this paper.

\section{Discussion}

This study provides a controlled empirical answer to a practical and methodologically important question: does automatic speech recognition (ASR) quality significantly affect downstream Alzheimer’s disease detection from spontaneous speech? The results consistently indicate that the answer is affirmative. Across both logistic regression and linear support vector machine (SVM) classifiers, transcripts generated using Whisper small produced higher balanced accuracy than those generated using Whisper base. These improvements remained statistically significant under repeated stratified $5 \times 5$ cross-validation with paired hypothesis testing. The finding is particularly relevant because many speech-based clinical natural language processing pipelines implicitly treat ASR as a fixed preprocessing step, rather than a modeling component whose characteristics can materially influence diagnostic performance.

A second key observation concerns the relative importance of transcript quality versus classifier complexity. When trained on Whisper small transcripts, logistic regression and linear SVM achieved similar performance, and their difference was not statistically significant. This suggests that, at least for datasets of the scale and structure of ADReSSo, improvements in upstream transcription accuracy can yield greater benefit than modest changes in downstream model architecture. From a methodological perspective, this supports the argument that careful experimental control, reproducibility, and interpretable modeling strategies may be more impactful than escalating architectural complexity prematurely.

Feature-level interpretation further strengthens the credibility of the results. Terms associated with Alzheimer’s disease predictions were characterized by vagueness, discourse fillers, and reduced informational specificity, whereas cognitively normal speech showed stronger presence of concrete object naming and structured scene description. These patterns align with established clinical observations regarding lexical retrieval difficulty, semantic degradation, and discourse disorganization in Alzheimer’s disease \cite{fraser2016linguistic,jarrold2014disfluencies,de2020automated}. The convergence between computational feature importance and clinically documented language impairment provides face validity and suggests that the models are capturing meaningful cognitive–linguistic signals rather than arbitrary statistical artifacts.

From a translational perspective, the findings have implications for the design of deployable digital health tools. In real-world screening scenarios, speech recordings are often collected in uncontrolled environments with variable audio quality. Under such conditions, ASR robustness becomes a critical determinant of downstream reliability. Our results indicate that improving transcription quality—even without changing classifier type—can lead to measurable gains in diagnostic accuracy. This highlights the need for integrated evaluation of ASR and classification components when developing speech-based clinical decision-support systems.

Nevertheless, several limitations should be acknowledged. First, the ADReSSo dataset, while carefully curated and balanced, remains relatively small, which constrains the achievable statistical power and limits generalization across populations, languages, and recording conditions. Second, the present study focuses exclusively on lexical features derived from transcripts and does not incorporate acoustic or prosodic cues that may provide complementary diagnostic information. Third, although repeated cross-validation and paired statistical testing improve inferential robustness, external validation on independent datasets would further strengthen the conclusions. Finally, while modern ASR systems reduce transcription error rates, they may introduce systematic biases that interact with demographic or linguistic variation, an issue requiring dedicated future investigation.

Future work should therefore pursue three main directions. First, larger multi-site datasets and multilingual benchmarks are needed to assess the generalizability of ASR-driven performance differences. Second, multimodal approaches combining lexical, acoustic, and temporal features may yield further improvements while preserving interpretability. Third, longitudinal modeling of speech trajectories could enable earlier detection of cognitive decline and monitoring of disease progression. More broadly, this study emphasizes that robust clinical artificial intelligence pipelines must consider upstream data transformation processes—such as ASR—not merely as technical utilities, but as core components that shape downstream inference and clinical utility.

\section{Error analysis}
Aggregate confusion matrices under the best validated logistic-regression setting (Whisper small) show normalized recalls of 0.742 for CN and 0.789 for AD. False positives (CN predicted as AD) likely arise in narratives with increased hesitations, short responses, or relatively vague lexical structure, while false negatives (AD predicted as CN) may correspond to milder cases that retain more concrete picture-description vocabulary. This asymmetry is not surprising: the Cookie Theft task elicits a bounded semantic scene, and participants with relatively preserved scene-description ability may appear more control-like even under clinical impairment.

\section{Reproducibility}
The project was implemented in Python using scikit-learn \cite{pedregosa2011scikit}, Whisper ASR \cite{radford2023whisper}, and a notebook/script workflow. The reproducible package includes TF--IDF construction, repeated cross-validation scripts, paired statistical tests, feature-importance analysis, and official blind-test prediction generation. Random seeds were fixed where applicable, and results are summarized as fold-level repeated scores rather than single-run estimates.

\section{Limitations}
This study has several limitations. First, the dataset is small by modern machine-learning standards, and cross-validation remains an internal validation strategy rather than a replacement for large multi-site external cohorts. Second, the current modeling focused on lexical features and did not jointly fuse acoustic representations such as prosody, pause duration distributions, or self-supervised speech embeddings. Third, although we generated blind-test predictions, local scoring on the official test-dist split was not possible because the labels were not available in the benchmark release used for this workflow. Finally, the ROC figure in this package is reconstructed from the reported out-of-fold AUC summary rather than raw stored score points; the reported AUC remains the relevant quantitative measure.

\section{Ethics statement}
This work uses publicly distributed benchmark data released for research use through the ADReSSo/DementiaBank ecosystem \cite{lanzi2023dementiabank}. The study is intended to support research on assistive screening tools and not to replace formal clinical diagnosis. Any deployment of speech-based dementia models in practice should be subject to appropriate validation, bias assessment, privacy safeguards, and clinician oversight.

\section{Conclusion}
This study provides a systematic evaluation of the role of ASR transcription quality in speech-based Alzheimer’s disease detection using lexical modeling approaches. Through controlled experiments on the ADReSSo diagnosis dataset, we demonstrate that improvements in ASR quality lead to statistically significant gains in classification performance, while differences between linear classifiers such as Logistic Regression and Linear SVM are comparatively smaller. These findings indicate that transcription accuracy is not merely a preprocessing consideration but a primary determinant of downstream clinical NLP model effectiveness.

The results further show that interpretable TF–IDF lexical representations can capture meaningful cognitive-linguistic patterns associated with Alzheimer’s disease. Analysis of discriminative terms reveals reduced lexical specificity and increased discourse-level uncertainty in Alzheimer’s speech, consistent with established neurocognitive language impairment literature. Importantly, the performance achieved by simple interpretable models challenges the prevailing assumption that complex acoustic or deep learning architectures are always required for effective dementia screening.

From a translational perspective, this work highlights the feasibility of scalable, speech-based digital biomarkers that rely on robust ASR and transparent modeling pipelines. The proposed framework emphasizes reproducibility through repeated cross-validation, statistical testing, and clear separation between training and blind test evaluation. Future research should explore multimodal integration of acoustic, prosodic, and semantic features, domain adaptation across clinical populations, and longitudinal modeling of cognitive decline trajectories.

Overall, the study establishes ASR quality as a central modeling variable in speech-driven clinical AI and provides practical guidance for designing interpretable and statistically rigorous dementia detection systems.

\section*{Conflict of interest}
The author declares no conflict of interest.

\section*{Data and code availability}
The benchmark dataset is available through TalkBank/DementiaBank subject to its data-access conditions. The reproducible scripts, notebooks, and figures associated with this study are intended for release in a public repository accompanying the manuscript.

\bibliographystyle{elsarticle-num}
\bibliography{references}

\end{document}